\documentclass[aps,prl,twocolumn,superscriptaddress,showpacs,floatfix,longbibliography]{revtex4-1}
\usepackage{amsmath,amssymb,amsfonts,float,graphics,epsfig,epstopdf,color,verbatim,tabularx,bm,multirow,appendix,hyperref}
\usepackage{lmodern}
\usepackage{color}
\usepackage{bm}

\DeclareMathOperator{\Tr}{Tr}

\begin{document}
	\title{Stable computation of entanglement entropy for 2D interacting fermion systems}
	
\author{Gaopei Pan}
\email{gppan@iphy.ac.cn}
\affiliation{Beijing National Laboratory for Condensed Matter Physics and Institute of Physics, Chinese Academy of Sciences, Beijing 100190, China}
\affiliation{School of Physical Sciences, University of Chinese Academy of Sciences, Beijing 100049, China}

\author{Yuan Da Liao}
\email{ydliao@fudan.edu.cn}
\affiliation{State Key Laboratory of Surface Physics, Fudan University, Shanghai 200438, China}
\affiliation{Center for Field Theory and Particle Physics, Department of Physics, Fudan University, Shanghai 200433, China}

\author{Weilun Jiang}
\affiliation{Beijing National Laboratory for Condensed Matter Physics and Institute of Physics, Chinese Academy of Sciences, Beijing 100190, China}
\affiliation{School of Physical Sciences, University of Chinese Academy of Sciences, Beijing 100049, China}

\author{Jonathan D'Emidio}
\affiliation{Donostia International Physics Center, P. Manuel de Lardizabal 4, 20018 Donostia-San Sebasti\'an, Spain}

\author{Yang Qi}
\affiliation{State Key Laboratory of Surface Physics, Fudan University, Shanghai 200438, China}
\affiliation{Center for Field Theory and Particle Physics, Department of Physics, Fudan University, Shanghai 200433, China}
\affiliation{Collaborative Innovation Center of Advanced Microstructures, Nanjing 210093, China}

\author{Zi Yang Meng}
\email{zymeng@hku.hk}
\affiliation{Department of Physics and HKU-UCAS Joint Institute
	of Theoretical and Computational Physics, The University of Hong Kong,
	Pokfulam Road, Hong Kong SAR, China}

\begin{abstract}
There is no doubt that the information hidden in entanglement entropy (EE), for example, the $n$-th order R\'enyi EE, i.e., $S^{A}_n=\frac{1}{1-n}\ln \Tr (\rho_A^n)$ where $\rho_A=\mathrm{Tr}_{\overline{A}}\rho$ is the reduced density matrix, can be used to infer the organizing principle of 2D interacting fermion systems, ranging from spontaneous symmetry breaking phases, quantum critical points to topologically ordered states. It is far from clear, however, whether the EE can actually be obtained with the precision required to observe these fundamental features -- usually in the form of universal finite size scaling behavior. Even for the prototypical 2D interacting fermion model -- the Hubbard model, to all existing numerical algorithms, the computation of the EE has not been succeeded with reliable data that the universal scaling regime can be accessed. Here we explain the reason for these unsuccessful attempts in EE computations in quantum Monte Carlo simulations in the past decades and more importantly, show how to overcome the conceptual and computational barrier with the incremental algorithm, such that the stable computation of the EE in 2D interacting fermion systems  can be achieved and universal scaling information can be extracted. Relevance towards the experimental 2D interacting fermion systems is discussed.
\end{abstract}

\date{\today}
\maketitle

\noindent{\textcolor{blue}{\it Introduction.}---} Entanglement witnesses can reveal the fundamental organizing principle of quantum many-body systems~\cite{cardyFinite1988,Srednicki1993,Christoph1994,calabreseEntanglement2004,fradkinEntanglement2006,Casini2006,Kitaev2006top,Levin2006,casini2007:EE2plus1,li2008entanglement,FSong2012,yanUnlocking2023,laflorencieQuantum2016,chandranHow2014,jiangFermion2022,liuFermion2022,poiblanceEntanglement2010,wangScaling2021,chenTopological2022,wangScaling2022,song2023deconfined,song2023quantum}. One of such witness is the entanglement entropy (EE), i.e. the $n$-th order R\'enyi EE $S^{A}_n=\frac{1}{1-n}\ln \Tr (\rho_A^n)$ where $\rho_A=\mathrm{Tr}_{\overline{A}}\rho$ is the reduced density matrix of a many-body Hamiltonian~\cite{calabreseEntanglement2004,fradkinEntanglement2006,Casini2006,Kitaev2006top,Levin2006,casini2007:EE2plus1,groverEntanglement2013,assaadEntanglement2014,Chang2014:EEhubbardbilayer,laflorencieQuantum2016,albaOut2017,parisenEntanglement2018,dEmidioEntanglement2020,zhaoScaling2022,zhaoMeasuring2022,dEmidioUniversal2022,liaoTeaching2023,swingle2010entanglement,hellingSpecial2010,cramer2007statistics,barthel2006entanglement,mishmashEntanglement2016}. The EE is an important quantity for the investigations of 2D and higher dimensional interacting fermion systems, as it can reveal the fundamental conformal field theory (CFT) data for the fermionic quantum critical points~\cite{cardyFinite1988,calabreseEntanglement2004,fradkinEntanglement2006,Casini2006}, the nature of the low-energy collective modes~\cite{swingle2010entanglement,groverEntanglement2013,assaadEntanglement2014,Chang2014:EEhubbardbilayer,laflorencieQuantum2016,dEmidioUniversal2022,liaoTeaching2023,barthel2006entanglement,mishmashEntanglement2016} and the topological information~\cite{Kitaev2006top,Levin2006}, which are usually difficult to compute otherwise. Therefore, to be able to compute the scaling behavior of the EE for 2D interacting fermion systems hold the key for understanding of properties of non-Fermi-liquid and strange metal states in the high-temperature superconductivity, the novel phases in quantum moir\'e materials, the fermion quantum criticalities and topological ordered states, etc. However, as we will explain below, the stable computation of EE for 2D interacting fermion systems have not been succeeded despite many attempts over the past decades. 

The EE of free fermion systems can be derived via the Widom-Sobolev formula~\cite{gioev2006entanglement,
	leschkeScaling2014,sobolevSchatten2014,
	sobolevWiener2015,swingle2010entanglement,
	hellingSpecial2010,jiangFermion2022} and results in the $L\log(L)$ scaling of a free Fermi surface in 2D~\cite{swingle2010entanglement,hellingSpecial2010,cramer2007statistics,barthel2006entanglement,mishmashEntanglement2016,jiangFermion2022}. The universal log-coefficient beyond the area law scaling for free Dirac fermions has also been obtained~\cite{casini2007:EE2plus1,Sahoo2016:RenyiQCP,helmesUniversal2016,jiangFermion2022,liuFermion2022}. The computation of the EE for interacting fermion lattice models in 2D has not been successful, with notable exceptions including topological EE computed from fractional quantum Hall groundstates~\cite{Zaletel2013:FQH,Zhu2015:NAFQH}. 

Since the computation of EE in 2D interacting fermion lattice models requires access of many-body wavefunction or partition function~\cite{calabreseEntanglement2004,fradkinEntanglement2006,Casini2006,Kitaev2006top,Levin2006,casini2007:EE2plus1}, the auxiliary-field determinant quantum Monte Carlo (DQMC) method becomes a good tool to possibly obtain the EE in the exponentially large Hibert space~\cite{groverEntanglement2013,assaadEntanglement2014,
Chang2014:EEhubbardbilayer,parisenEntanglement2018,peterRenyi2014,
assaadStable2015,broeckerNumerical2016,dEmidioUniversal2022,liaoTeaching2023,daliaoControllable2023}. In the past decades, significant algorithmic advances in the computation of R\'enyi EE have been made. This was spearheaded by the original work of Grover~\cite{groverEntanglement2013} who used the free fermion decomposition of the reduced density matrix to identify an estimator based on independent auxiliary-field configurations.  Despite its elegance, early implementations of this approach revealed severe statistical errors at strong coupling and not-even-large subsystem sizes~\cite{assaadEntanglement2014,Chang2014:EEhubbardbilayer}. This motivated translating the highly successful replica approach from path-integral spin systems~\cite{hastings2010measuring} into the auxiliary-field fermion language \cite{peterRenyi2014,assaadStable2015,broeckerNumerical2016}, which however proved cumbersome since it required introducing a replicated environment for the entangling subsystem and using an imaginary-time dependent Hamiltonian, thus substantially increasing the computation burden (the computational complexity of DQMC scales as $O(\beta N^3)$ with $\beta=\frac{1}{T}$ the inverse temperature and $N=L^d$ for $d$ spatial dimension systems with linear size $L$).  Furthermore it suffered from subtle stability issues regarding the computation of Green's functions.  
In the end, all attempts thus far in computing R\'enyi EE for interacting fermions in 2D have not achieved the precision required to extract, in the simplest square lattice Hubbard model, an area law plus universal log due to Goldstone modes~\cite{metlitskiEntanglement2011}.

\begin{figure}[htp!]
	\includegraphics[width=\columnwidth]{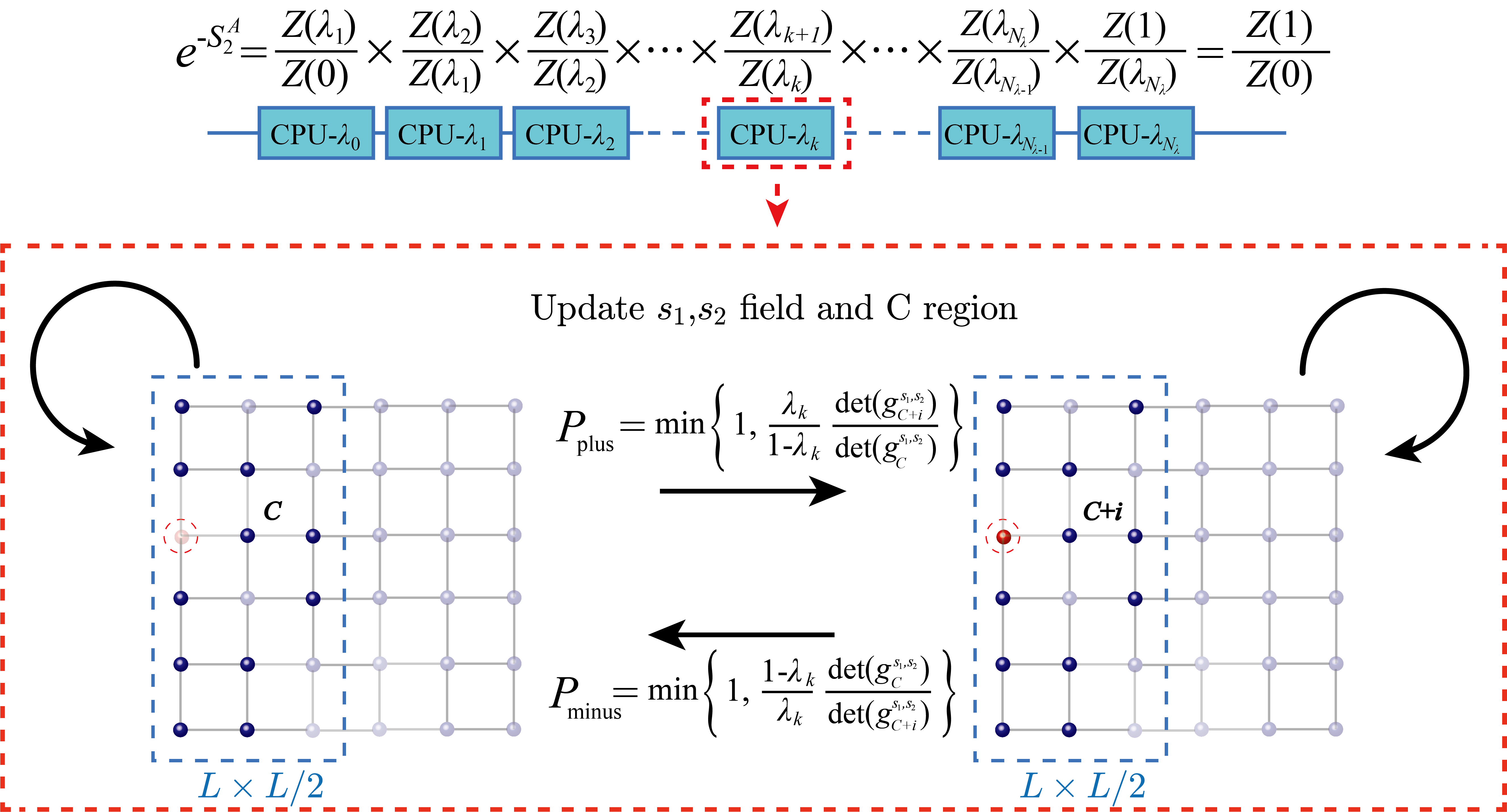}
	\caption{
		\textbf{The incremental computation of EE.}  The entanglement region $A$ is denoted as the blue dashed box. Accroding to Eq.~\eqref{eq:eq5}, we split the computation into the parallel execution of many ratios, where each ratio is bounded with a scale of unity and the sites inside $A$ for each parallel piece are not fixed but changes stochastically. The adding and removing site (denoted as the red dot and dashed circle) in $A$ are carried out with probabilities $P_{\text{plus}}$ and $P_{\text{minus}}$. The QMC updates of the auxiliary field for each parallel piece are carried out independently. When update the $s_1,s_2$ field, the sites in $A$ and in the environment are fixed, as denoted by the circular arrows.}
	\label{fig:fig1}
\end{figure}  

On the other hand, since the QMC computational complexity in spin/boson systems scales as $O(\beta N)$, the EE of 2D Heisenberg models~\cite{kallin2011anomalies,hastings2010measuring,humeniuk2012quantum,helmes2014entanglement,kulchytskyy2015detecting} and other related systems~\cite{isakovTopological2011,laflorencieQuantum2016} have had much success, although the data quality is always a serious issue for extracting the expected universal scaling coefficients. This problem is completely solved by the introduction of the {\it incremental} algorithm~\cite{albaOut2017,dEmidioEntanglement2020,zhaoMeasuring2022,zhaoScaling2022}. The algorithm converts the computation of the R\'enyi EE into the parallel execution of the
Jarzynski equality~\cite{Jarzynski1997} of the free energy difference between partition functions on replicated manifolds, making the precise determination of EE scaling on various 2D quantum spin models possible with exquisite data quality. By using the algorithm, controlled results with the expected CFT information can then be obtained, including in the N\'eel phase of antiferromagnetic Heisenberg model, at the (2+1)D O(3) quantum critical point, the deconfined quantum critical point and inside the $Z_2$ topological ordered Kagome quantum spin liquid~\cite{dEmidioEntanglement2020,zhaoMeasuring2022,zhaoScaling2022}, to name a few.

It is in the process of developing the {\it incremental} algorithm into DQMC for the entanglement computation in interacting fermion systems~\cite{dEmidioUniversal2022}, that we understand the reason why the previous {\it direct} implementation of the EE computation~\cite{groverEntanglement2013,assaadEntanglement2014,Chang2014:EEhubbardbilayer,assaadStable2015,broeckerNumerical2016,peterRenyi2014,toldinFntanglement2018} does not work -- not because of the heavy computation added to the already expensive DQMC by adding replicas, but because the direct computation does not use the correct sampling weight to construct a proper Markov chain Monte Carlo simulation. The incremental algorithm~\cite{dEmidioUniversal2022,liaoTeaching2023}, on the other hand, features two key improvements:
\begin{enumerate}
\item designing the effective 
 Monte Carlo importance sampling weight for EE computations and 
\item conditioning the exponentially small partition function ratio into a parallel execution of values with scales of unity
\end{enumerate} 
and consequently offers the correct scheme that can be utilized to extract the EE with reliable data quality and controllable polynomial computation complexity. Here, we use the prototypical example of 2D interacting fermion lattice models -- the square lattice Hubbard model --  to fully explain the simple but fundamental breakthrough of the {\it incremental} over the previous {\it direct} computation of EE. The algorithm opens the avenue for the stable EE computation for 2D fermion quantum matter and can be used to extract the universal information for the quantum critical metal and non-Fermi-liquid~\cite{liaoDiracI2022,liaoDiracII2022,liaoDiracIII2022,xuRevealing2019,jiangFermion2022,liuItinerant2019,panSport2022,xuNonFermi2017,patelUniversal2022,esterlisLarge2021,luntsNon2022}, the fermion deconfined quantum critical point~\cite{liaoTeaching2023,liuFermion2022,christosModel2023,liuSuperconductivity2019}, the correlated flat-band Moir\'e materials~\cite{daliaoValence2019,daliaoCorrelation,liaoCorrelated2021,panThermodynamic2023,zhangQuantum2023,huangEvolution2023} and kagome metals~\cite{yinTopological2022,kangDirac2020,sankarObservation2023} and the entanglement spectra and Hamiltonian~\cite{li2008entanglement,chandranHow2014,poiblanceEntanglement2010,yanUnlocking2023,songReversing2022,assaadEntanglement2014,assaadStable2015}, which cannot be accessed with other methods.

\begin{figure}[htp!]
\includegraphics[width=0.8\columnwidth]{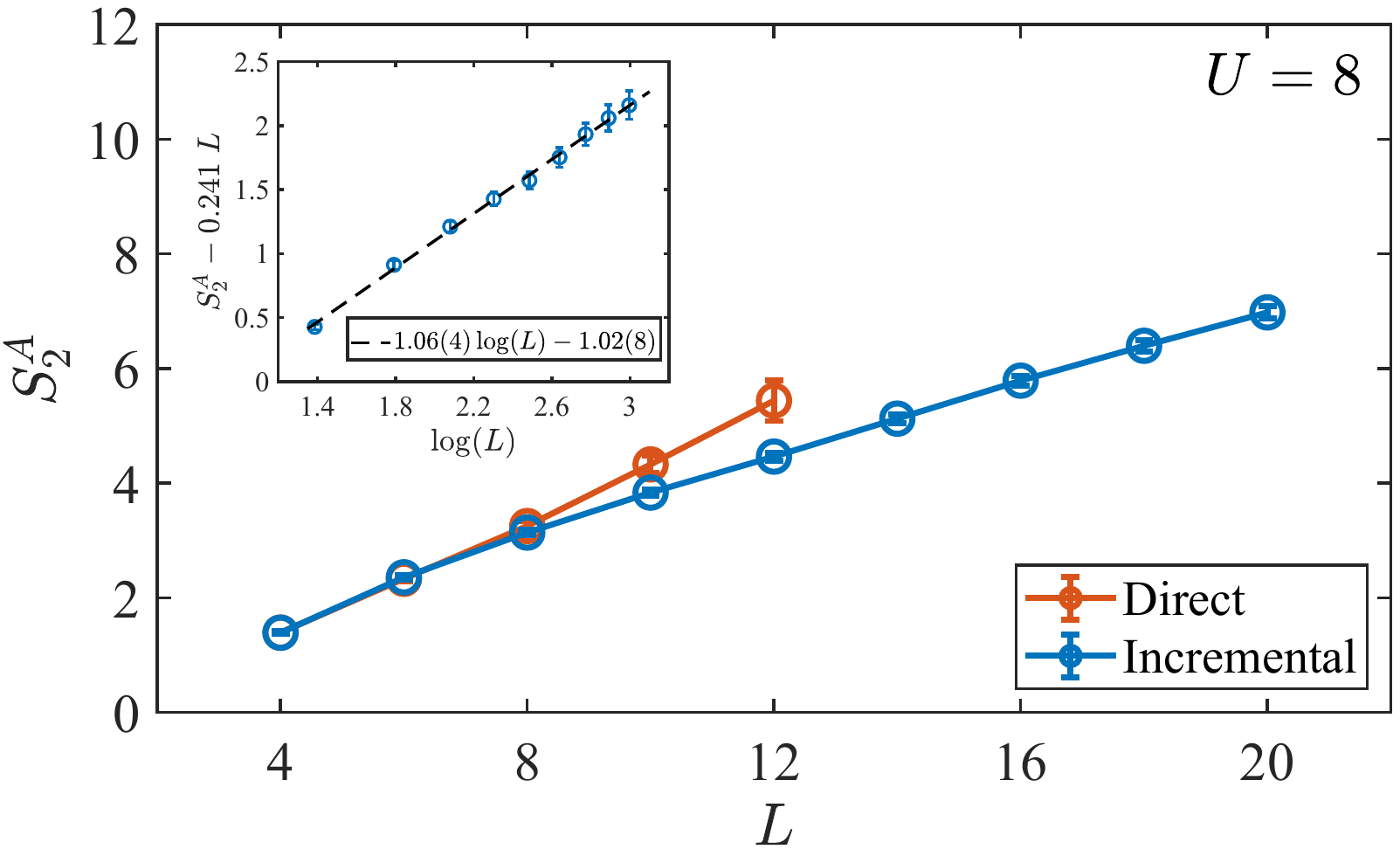}
\caption{\textbf{The EE of square lattice Hubbard model at $U=8$.} The entanglement region $A$ is of size $L\times L/2$. The red and blue lines are from the direct and incremental methods, respectively. The deviation of the direct computation for $L\ge8$ is obvious. The inset shows the incremental data of $S^{A}_2 - 0.241 L$ versus $\log(L)$, with the slope (denoted as the black dashed line) representing the universal log-coefficient $s=\frac{N_G}{2}=1$ in Eq.~\eqref{eq:eq4}, our fitting results of $s=1.06(4)$ is fully consistent with expected behavior of N\'eel antiferromagnetic Mott insulator with $N_G=2$. It is important to note that the errorbars of direct data(red dots) are unestimated, as the mean values haven't converged shown in Fig.~\ref{fig:fig4}.}
\label{fig:fig2}
\end{figure}

\noindent{\textcolor{blue}{\it Model and Method.}---}
We study the 2nd R\'enyi EE for the square lattice Hubbard model at half-filling, with the Hamiltonian $H=-t\sum_{\langle i,j \rangle} (c^{\dagger}_{i,\sigma}c_{j,\sigma} + h.c.) + \frac{U}{2}\sum_{i} (n_{i,\uparrow}+n_{i,\downarrow}-1)^2$, where $c^{\dagger}_{i,\sigma}$ and $c_{i,\sigma}$ denote the creation and annihilation operators with spin $\sigma=\uparrow,\downarrow$ on site $i$, $\langle i,j \rangle$ represents the nearest neighbor hopping, $n_{i,\sigma}= c^{\dagger}_{i,\sigma} c_{i,\sigma} $ is the particle number operator for spin $\sigma$, and $U/t$ tunes the onsite interaction strength, with $t=1$ the energy unit.

We utilize the projector DQMC method to compute the R\'enyi EE. As described in the Supplementary Material (SM)~\cite{suppl} and literature~\cite{assaadWorld-line2008,xuMonte2019,daliaoValence2019,daliaoCorrelation,liaoCorrelated2021,liaoDiracI2022,liaoDiracII2022,liaoDiracIII2022}, it carries out a Hubbard-Stratonovich transformation to introduce an auxiliary field $\{ s \}$ to decouple the quartic fermion interaction and compute ground-state observable as $\langle O \rangle = \frac{\sum_{\{s\}} W^s \langle O \rangle ^s}{\sum_{\{s\}} W^s}$, where $W^s$ is the unnormalized weight of configuration $s$ proportional to a determinant whose elements depends on $s$~\cite{suppl}. To calculate the R\'enyi EE of interacting fermions in DQMC, Grover introduced a {\it direct} formula~\cite{groverEntanglement2013} based on the free fermion decomposition of the reduced density matrix $\rho_A$ (with entangling region $A$) using two independent auxiliary field replicas, such that the 2nd R\'enyi EE $S_2^{A}$ is given by
  \begin{equation}
 	e^{-S_2^A}=\frac{Z(1)}{Z(0)}:=\frac{ \sum_{\{s_1,s_2\}}\mathbf{W}^{s_1,s_2}\det g_A^{s_1,s_2} }{\sum_{\{s_1,s_2\}} \mathbf{W}^{s_1,s_2}},
 	\label{eq:eq3}
 \end{equation}
where $\mathbf{W}^{s_1,s_2}=W^{s_1} W^{s_2}$, $g_A^{s_1,s_2}= G_A^{s_1} G_A^{s_2}+\left(\mathbb{I}-G_A^{s_1}\right)\left(\mathbb{I}-G_A^{s_2}\right) $ is the Grover matrix connecting the Green's functions $G^{s_1}$ and $G^{s_2}$ of the two replicas on $A$. $Z(1)$ stands for the partition function with the fully connected entangling region between the two replicas and $Z(0)$ the partition function of two independent replicas, we use $\lambda \in [0,1]$ to parametrize the evolution from $Z(\lambda=0)$ to $Z(\lambda=1)$.

Based on Eq.~\eqref{eq:eq3}, one can compute the $S_2^{A}$ as in conventional DQMC simulations with the configurational weights $\mathbf{W}^{s_1,s_2}$, and this is indeed what has been implemented in previous attempts~\cite{groverEntanglement2013,assaadEntanglement2014,Chang2014:EEhubbardbilayer}. But it was noticed that the obtained EE suffered greatly from numerical instability issue with poor data quality that they cannot be used to extract the universal scaling behavior 
\begin{equation}
S_2^{A}(L) = a L\log L + b L + s \log L + c
\label{eq:eq4}
\end{equation}
where the coefficients $a$ stems from the 2D Fermi surface and can be derived at the non-interacting limit~\cite{gioev2006entanglement,swingle2010entanglement,hellingSpecial2010,cramer2007statistics,barthel2006entanglement,mishmashEntanglement2016,jiangFermion2022} (see Eq.(S7) and Fig. S1), $b$ governs the area law scaling and $s$ is the universal corner contributions at critical points~\cite{fradkinEntanglement2006,casini2007:EE2plus1,laflorencieQuantum2016}, or is proportional to the number of Goldstone modes in symmetry broken phases~\cite{metlitskiEntanglement2011}. For example, for 2D Hubbard model at $U=8$, the $s=\frac{N_G}{2}=1$ where $N_G=2$ is the number of Goldstone modes for a N\'eel state (see Fig.~\ref{fig:fig2} below), a result that has eluded implementations using the direct approach of Grover.

\begin{figure}[htp!]
\includegraphics[width=0.8\columnwidth]{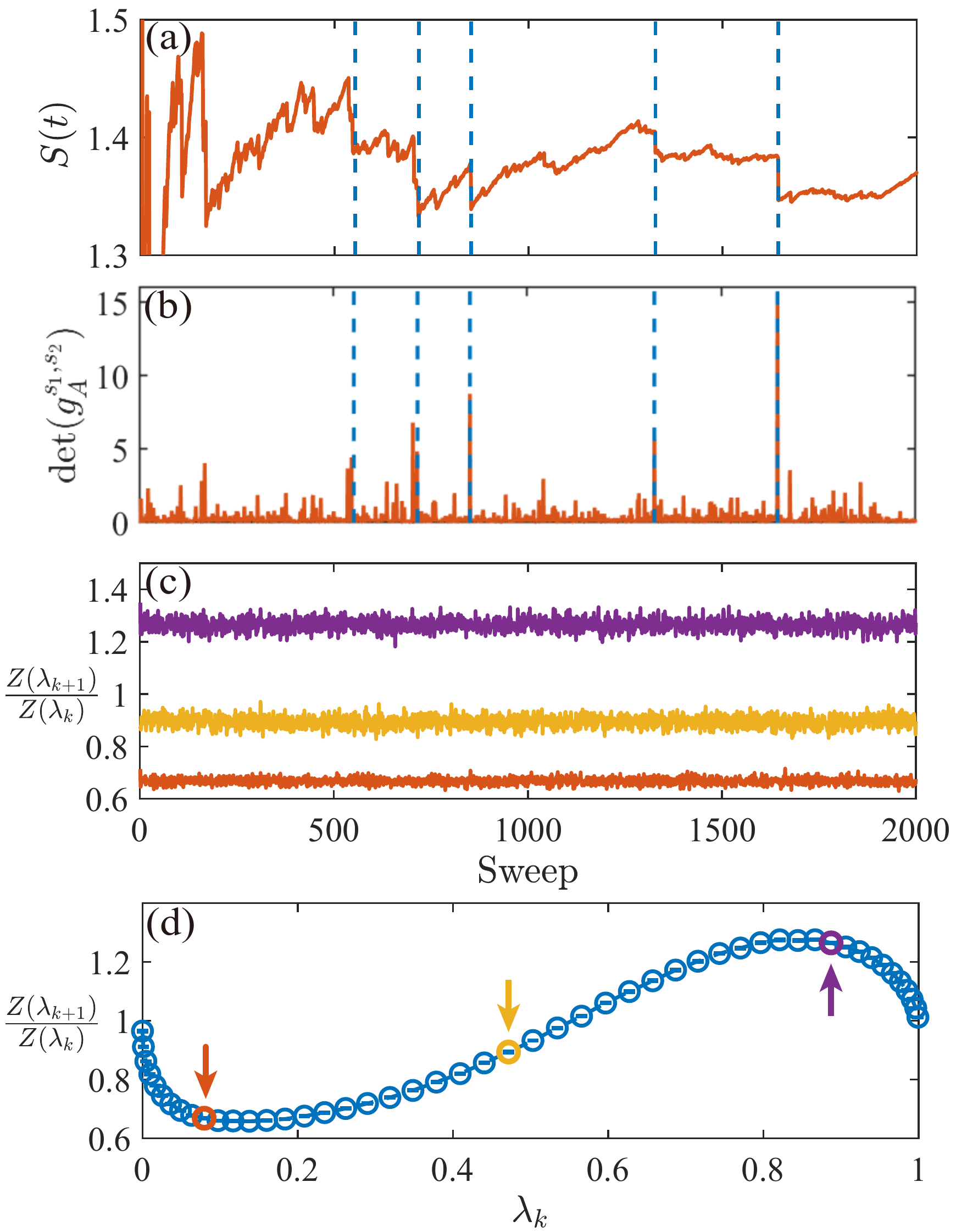}
\caption{\textbf{Difference between direct and incremental measurements.} (a) and (b) show the time series of $\det(g^{s_1,s_2}_A)$ and $S(t)$ from a single Markov chain using direct method with $U=8, L=4$. Both observables are clearly not normal distributed and the rare events in the form of the sudden drop in  (a) and spikes in (b), denoted by blue dotted lines, are clearly seen. (c) and (d) $\frac{Z(\lambda_{k+1})}{Z(\lambda_k)}$ for $U=8, L=10$ by incremental method with $\lambda\in[0,1]$. The number of $\lambda_k$, $N_{\lambda}$ is 50 and each piece has the value of scale unity. The three arrows in panel(d) point out the position of three different $\lambda_k$ values whose time series are shown on in panel(c). The observables are now normal distributed.}
\label{fig:fig3}
\end{figure}  

What has been seen, however, is that for slightly larger system sizes $(L\ge 8)$ and slightly stronger interactions $(U\ge4)$, the distribution of the Grover determinants became very broad with spikes (rare events). We find if one insists on direct computation of EE with Eq.~\eqref{eq:eq3}, it is these rare events that actually make great contributions to the expectation value of EE, but since they occur less often with respect to the $L$ and $U$, one will certainly run into problem with increased $L$ and $U$. This means the {\it direct} computation of EE in Eq.~\eqref{eq:eq3} follows the incorrect distribution $\mathbf{W}^{s_1,s_2}$, and consequently does not average according to the important sampling of a Markov chain Monte Carlo process.

To address this issue, i.e. to sample properly in the replicated configurational space of the EE computation, the incremental algorithm for fermions was recently developed in Ref.~\cite{dEmidioUniversal2022} and further applied in Ref.~\cite{liaoTeaching2023}. As sketched in Fig.~\ref{fig:fig1}, the incremental algorithm has improved the direct computation in two main points:

First, it introduces a new auxiliary sampling configuration, namely the subset $C$ of the entanglement region $A$, which, instead of calculating $e^{-S_2^A}$ directly, converts the computation of $e^{-S_2^A}$ into a parallel execution of incremental ratios as
\begin{equation}
 	e^{-S_2^A}=\frac{Z(1)}{Z(0)}:=\frac{Z(\lambda_1)}{Z(0)}\frac{Z(\lambda_2)}{Z(\lambda_1)}\cdots\frac{Z(\lambda_{k+1})}{Z(\lambda_k)} \cdots\frac{Z(1)}{Z(\lambda_{N_\lambda})},
\label{eq:eq5}
\end{equation}
where $Z(\lambda)=\sum_{C \subseteq A} \lambda^{N_C}(1-\lambda)^{N_A-N_C} Z_C$~\cite{dEmidioEntanglement2020,dEmidioUniversal2022} with $\lambda \in[0,1]$, $N_C$ ($N_A$) is the number of site in region $C$ $(A)$ and $Z_C=\sum_{\left\{s_1\right\},\left\{s_2\right\}} \mathbf{W}^{s_1,s_2} \operatorname{det} g_C^{s_1, s_2}$. $N_\lambda$ is the number of $\lambda_k$. $\frac{Z(\lambda_{k+1})}{Z(\lambda_k)}$ is computed as
\begin{equation}
\frac{Z\left(\lambda_{k+1}\right)}{Z\left(\lambda_k\right)}=\frac{\sum_{\{s_1, s_2, C \subseteq A\}}  \mathbf{W}^{s_1,s_2}_{C}(\lambda_k) O_C\left(\lambda_k,\lambda_{k+1}\right) }{\sum_{\{s_1, s_2, C \subseteq A\}} \mathbf{W}^{s_1,s_2}_{C}(\lambda_k) },
\label{eq:eq6}
\end{equation}
where the observable for the EE is $O_C\left(\lambda_k,\lambda_{k+1}\right)=\left(\frac{\lambda_{k+1}}{\lambda_k}\right)^{N_C}\left(\frac{1-\lambda_{k+1}}{1-\lambda_k}\right)^{N_A-N_C}$, and the sampling weight of the EE computation is
\begin{equation}
\mathbf{W}^{s_1,s_2}_{C}(\lambda_k) = \lambda_k^{N_C}\left(1-\lambda_k\right)^{N_A-N_C} \mathbf{W}^{s_1,s_2} \operatorname{det} g_C^{s_1, s_2}.
\label{eq:eq8}
\end{equation}
We note the weight ratio $ \frac{\mathbf{W}^{s_1^{\prime},s_2}_{C}(\lambda_k)}{\mathbf{W}^{s_1,s_2}_{C}(\lambda_k)} = \frac{W^{s_1^{\prime}}\det g_C^{s_1^{\prime},s_2} }{ W^{s_1} \det g_C^{s_1,s_2}}$ of incremental sampling in Eq.~\eqref{eq:eq6} is explicitly different from that of direct sampling $ \frac{\mathbf{W}^{s_1^{\prime},s_2}}{\mathbf{W}^{s_1,s_2}} = \frac{W^{s_1^{\prime}}}{W^{s_1} }$ in Eq.~\eqref{eq:eq3}, in that it contains the contribution from determinant of Grover matrix. In addition, the incremental method updates the configurations $C$ stochastically with probabilities
$P_{\text{plus}}$ %=\frac{\lambda_k}{1-\lambda_k} \frac{\operatorname{det}\left(g_{C+i}^{s_1, s_2}\right)}{\operatorname{det}\left(g_C^{S_1, s_2}\right)}$ 
and $P_{\text{minus}}$ %=\frac{1-\lambda_k}{\lambda_k} \frac{\operatorname{det}\left(g_{C-i}^{s_1, s_2}\right)}{\operatorname{det}\left(g_{C}^{s_1, s_2}\right)}$ 
for adding or moving one site from region $C$, as shown in Fig.~\ref{fig:fig1}. When sampling according to Eq.~\eqref{eq:eq8}, there will be no spikes in the observable $O_{C}(\lambda_k,\lambda_{k+1})$, provided $\lambda_k$ and $\lambda_{k+1}$ are close enough such that their histograms of $N_C$ overlap.  The ensemble average can then be properly carried out.

Second, we find each piece of the ratio $\frac{Z\left(\lambda_{k+1}\right)}{Z\left(\lambda_k\right)}$ in Eq.~\eqref{eq:eq5} can be computed independently, which means massive parallelization of the high-performance computation (denoted in Fig.~\ref{fig:fig1}) can greatly reduce the computation time. Although the $e^{-S^{A}_2}$ is eventually an exponentially small quantity, each piece in the product of Eq.~\eqref{eq:eq5} actually has well-behaved bound of the scale of unity, since the independent computation of the $\frac{Z\left(\lambda_{k+1}\right)}{Z\left(\lambda_k\right)}$ is very well-behaved, their product gives rise to the controlled EE with the same polynomial complexity as DQMC. The increments $\frac{Z\left(\lambda_{k+1}\right)}{Z\left(\lambda_k\right)}$ of $O(1)$ and their histograms in the non-interacting cases are shown in the SM~\cite{suppl} (see also references~\cite{pasqualeExact2012} therein).

\begin{figure}[htp!]
	\includegraphics[width=0.8\columnwidth]{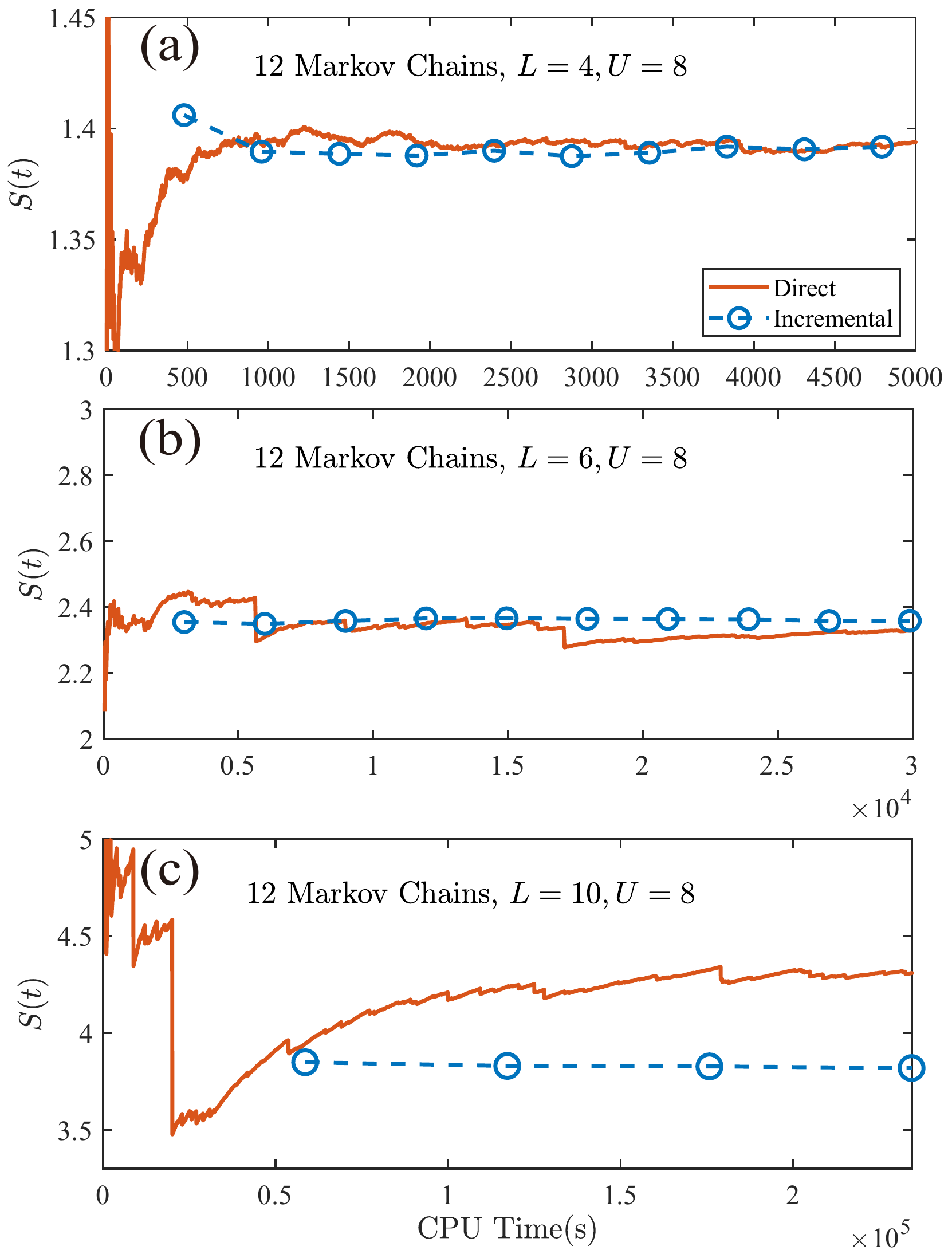}
	\caption{\textbf{Convergence comparison between the direct and the incremental methods.} (a) For $L=4,U=8$, the direct method (red line) can slowly converge to exact value while the blue dots, the incremental method converges fast. (b) For $L=6,U=8$, the direct method converge within a reasonable CPU time but with big fluctuations, the incremental method converges fast. (c) For $L=10,U=8$, the direct method cannot converge within the reasonable CPU time, the incremental method converges fast.}
	\label{fig:fig4}
\end{figure} 

\noindent{\textcolor{blue}{\it Results in 2D Hubbard model.}---}
We have carried out the EE computation for square lattice Hubbard model with $L=4,8,10,12,16,20$. Most of our data are obtained at $U=8$ where the system is in antiferromagnetic Mott insulator state. The $U=0$ limit is discussed in the SM~\cite{suppl}, where the computed $aL\log L$ with the coefficient $a=\frac{1}{2}$ in Eq.~\eqref{eq:eq4} obtained in full agreement with the analytic expectation from the Widom-Sobolev formula~\cite{gioev2006entanglement,leschkeScaling2014,sobolevSchatten2014,sobolevWiener2015,swingle2010entanglement,jiangFermion2022}. 

The results of EE at $U=8$ are shown in Fig.~\ref{fig:fig2}. Here the entanglement region is half of the lattice: $L/2 \times L$. One clearly sees that when the system size $L$ is small, the results of the two methods coincide, but when the size gradually increases, the mean value of the direct method starts to deviate from the expected behavior of the incremental one.

Moreover, since the half-filled square lattice Hubbard is always in an antiferromagnetic Mott insulating phase ($U>0$), the $S_2^{A}$ of the system with spontaneous broken SU(2) continuous symmetry should have a form in Eq.~\eqref{eq:eq4} with $a=0$, $b$ finite and the universal log-coefficient $s=\frac{N_G}{2}=1$ where $N_G=2$ is the number of the Goldstone modes~\cite{metlitskiEntanglement2011,laflorencieQuantum2016}. As shown in the inset of Fig.~\ref{fig:fig2}, the log-coefficient after extracting the area law term is $1.06(4)$, well consistent with the theoretical expected value $1$. The results of the direct computation will not be able to perform such analysis.

To reveal the difference of the two methods, we record the time series of EE computation along the Markov chain, $S(t)=-\log\left(\frac{1}{t} \sum^t_{i=1} e^{-S_2^A}(i) \right)$, where $S(t)$ represents the expectation value of  observable $S_2^A$ after first $t$ DQMC sweeps. As shown in Fig.~\ref{fig:fig3}(b), for the direct method, $S_2^A$ does not follow normal distribution, and whenever a peak is sampled, there is an obvious drop in the mean value of EE calculated, as shown in Fig.~\ref{fig:fig3}(a). The $S(t)$ is affected by these rare events, which renders the direct computation with very poor performance. As $L$ and $U$ increases, a very long Markov chain is needed to obtain accurate values, as shown in Fig.~\ref{fig:fig3}(b) and red lines in Fig.~\ref{fig:fig4}. In fact, from Fig.~\ref{fig:fig4} (c), one sees for $L=10$ and $U=8$, the direct $S(t)$ has not converged. The incremental EE has no problem. For each parallel piece, the range of the observable is controlled as we have considered the determinant of the Grover matrix in the weight during sampling in Eq.~\eqref{eq:eq8}. In Fig.~\ref{fig:fig3}(d), the range of the partition function ratios is given, with $L=10,U=8$ and $\lambda_k=\left[\sin\frac{(0.002+50(k-1))\pi}{2}\right]^2$. And the sampling distributions of three colored points are shown in Fig.~\ref{fig:fig3}(c). The incremental method with its fast convergence and parallel computation, clearly outperform the direction computation.

\noindent{\textcolor{blue}{\it Discussion.}---} By utilizing the square lattice Hubbard model, we reveal the fundamental difference between the {\it direct} and {\it incremental} computation of EE in that, the two major improvements i). designing the effective Monte Carlo sampling weight and ii). conditioning the exponential factor of partition function ratios into parallel execution of values with scale of unity, bestow the incremental method the access of the entanglement scaling behavior of 2D interacting fermion models. Our approach establishes the paradigm of the EE computation for 2D highly entangled fermion quantum matter and probably has the potential to offer the universal experimentally measurable quantities %, such as the universal log-coefficients, 
to guide experiments in quantum critical metal and non-Fermi-liquid~\cite{liaoDiracI2022,liaoDiracII2022,liaoDiracIII2022,xuRevealing2019,jiangFermion2022,liuItinerant2019,panSport2022,xuNonFermi2017,patelUniversal2022,esterlisLarge2021,luntsNon2022}, the fermion deconfined quantum critical point~\cite{liaoTeaching2023,liuFermion2022,christosModel2023,liuSuperconductivity2019}, the correlated flat-band Moir\'e materials~\cite{panThermodynamic2023,zhangQuantum2023,huangEvolution2023,daliaoValence2019,daliaoCorrelation,liaoCorrelated2021} and kagome metals~\cite{yinTopological2022,kangDirac2020,sankarObservation2023} and the entanglement spectra and Hamiltonian in 2D interacting fermion systems~\cite{li2008entanglement,chandranHow2014,
poiblanceEntanglement2010,yanUnlocking2023,songReversing2022,
assaadEntanglement2014,assaadStable2015}.

\begin{acknowledgments}
{\it Acknowledgments}\,---\, 
We thank Jiarui Zhao, Zheng Yan for collaborations on incremental algorithm for spin/boson systems~\cite{zhaoMeasuring2022,zhaoScaling2022} and inspiring discussions on the related topic. We thank Fakher Assaad for bringing our attention to the instability issue of EE computation over the years. GPP, WLJ and ZYM acknowledge support from the RGC of Hong Kong SAR of China (Project Nos. 17301420, 17301721, AoE/P-701/20, 17309822, HKU C7037-22G), the ANR/RGC Joint Research Scheme sponsored by Research Grants Council of Hong Kong SAR of China and French National Reserach Agency (Project No. A\_HKU703/22). 
YDL acknowledges support from National Natural Science Foundation of China (Grant No. 12247114) and the China Postdoctoral Science Foundation (Grants Nos. 2021M700857 and  2021TQ0076).
\end{acknowledgments}

\bibliography{ref.bib}
\bibliographystyle{apsrev4-1}

\clearpage
\onecolumngrid

\begin{center}
\textbf{\large Supplementary Material for "Stable computation of entanglement entropy for 2D interacting fermion systems"}
\end{center}
%\appendix
\setcounter{equation}{0}
\setcounter{figure}{0}
\setcounter{table}{0}
\setcounter{page}{1}
\makeatletter
\renewcommand{\theequation}{S\arabic{equation}}
\renewcommand{\thefigure}{S\arabic{figure}}
\setcounter{secnumdepth}{3}

\section{More details of projector DQMC}
In this study, we focus on the calculation of the 2nd R\'enyi entanglement entropy $S_2^A$ for the square lattice Hubbard model with $N=L^2$ sites. As $S_2^A$ is a ground-state observable, the projector DQMC method is particularly suitable to compute this quantity. This method obtains the ground state $\vert \Psi_0 \rangle$ by projecting a trial wave function $\vert \Psi_T \rangle$ through a relation $\vert \Psi_0 \rangle = \lim\limits_{\Theta \to \infty} e^{-\Theta H} \vert \Psi_T \rangle$, where $\Theta$ represents the projection time and $H$ denotes the Hamiltonian of the system. And the physical observable $\hat{O}$ is given as
\begin{equation}
\label{eq:eqS1}
\langle \hat{O} \rangle = \frac{\langle \Psi_0 \vert \hat{O} \vert \Psi_0 \rangle}{\langle \Psi_0 \vert \Psi_0 \rangle} 
= \lim\limits_{\Theta \to \infty} \frac{\langle \Psi_T \vert  e^{-\Theta H } \hat{O}  e^{-\Theta H} \vert \Psi_T \rangle}{\langle \Psi_T \vert  e^{-2\Theta H} \vert \Psi_T \rangle} .
\end{equation}

The Hamiltonian $H$ consists of two parts: the non-interacting $H_0=-t \sum_{\langle i j\rangle, \sigma} \left(c_{i \sigma}^{\dagger} c_{j \sigma}+\text { H.c. }\right)$ and the interacting $H_U=U \sum_{i}\left(n_{i,\uparrow}+n_{i,\downarrow}-1\right)^2$ terms, which do not commute. 
We need to employ Trotter decomposition to discretize the projection length $2\Theta$ into $l_\tau$ imaginary time slices ($2\Theta=l_\tau \Delta_\tau$) and have 
\begin{equation}
\langle\Psi_{T}|e^{-2\Theta H}|\Psi_{T}\rangle=\langle\Psi_{T}|\left(e^{-\Delta_\tau H_{0}}e^{-\Delta_\tau H_{U}}\right)^{l_\tau}|\Psi_{T}\rangle+\mathcal{O}(\Delta{\tau}^{2}).
\label{eq:eqS2}
\end{equation}
It is worth to note that one should set a small value for the Trotter discretization parameter $\Delta_\tau$, as the Trotter decomposition process introduces a small systematic error proportional to $\Delta_{\tau}^2$.

To decouple the quartic fermionic term in $H_U$, we employ a SU(2) symmetric Hubbard-Stratonovich (HS) transformation at site $i$ and time slice $l_\tau$
\begin{equation}
e^{-\Delta_\tau U(n_{i,\uparrow}+n_{i,\downarrow}-1)^{2}}=\frac{1}{4}\sum_{\{s_{i,l_\tau}\}}\gamma(s_{i,l_\tau})e^{\alpha\eta(s_{i,l_\tau})\left(n_{i,\uparrow}+n_{i,\downarrow}-1\right)}
\label{eq:decompo}
\end{equation}
with $\alpha=\sqrt{-\Delta\tau U}$, $\gamma(\pm1)=1+\sqrt{6}/3$,
$\gamma(\pm2)=1-\sqrt{6}/3$, $\eta(\pm1)=\pm\sqrt{2(3-\sqrt{6})}$,
$\eta(\pm2)=\pm\sqrt{2(3+\sqrt{6})}$, which transforms the quartic term into a quadratic one.
Then, we have
   \begin{eqnarray}
\langle\Psi_{T}|e^{-2\Theta H}|\Psi_{T}\rangle=\sum_{\{s_{i,l_\tau}\}}\left[\left(\prod_{l_\tau}^{L_{\tau}}\prod_{i}^{N}\gamma(s_{i,l_\tau})e^{-\alpha\eta(s_{i,l_\tau})}\det\left[P^{\dagger}B^{s_{i,l_\tau}}(2\Theta,0)P\right]\right)\right]
\label{eq:mcweight}
   \end{eqnarray}
where $P$ is the coefficient matrix of trial wave function $|\Psi_T\rangle$; $B^{s}(2\Theta,0)$ is defined as
\begin{equation}
B^{s}(\tau_{2},\tau_{1})= \prod_{l_\tau=l_{1}+1}^{l_{2}} \left( \mathrm{e}^{-\Delta_\tau H_0} \prod_{i}^{N} \mathrm{e}^{\alpha\eta(s) \left( n_{i,\uparrow}+n_{i,\downarrow}\right)} \right)
\end{equation}
with $l_1 \Delta_\tau = \tau_1$ and $l_2 \Delta_\tau = \tau_2$, and has a property $B^{s}(\tau_3,\tau_1)=B^{s}(\tau_3,\tau_2)B^{s}(\tau_2,\tau_1)$.
With these notations, the unormalized weight $W^{s_{i,l_\tau}}$ of Eq.~(2) in the main text could be given explicitly as
\begin{equation}
W^{s_{i,l_\tau}} = \gamma(s_{i,l_\tau})e^{-\alpha\eta(s_{i,l_\tau})}\det\left[P^{\dagger}B^{s_{i,l_\tau}}(2\Theta,0)P\right].
\end{equation}
In practice, we choose the ground state wavefunction of $H_0$ with as the trial wave function. 
The measurements are performed near $\tau=\Theta$, we set projection time $2\Theta = L$, discrete time slice $\Delta_\tau=0.1$.

\section{Non-interacting limit}

\begin{figure}[htp!]
	\includegraphics[width=0.5\columnwidth]{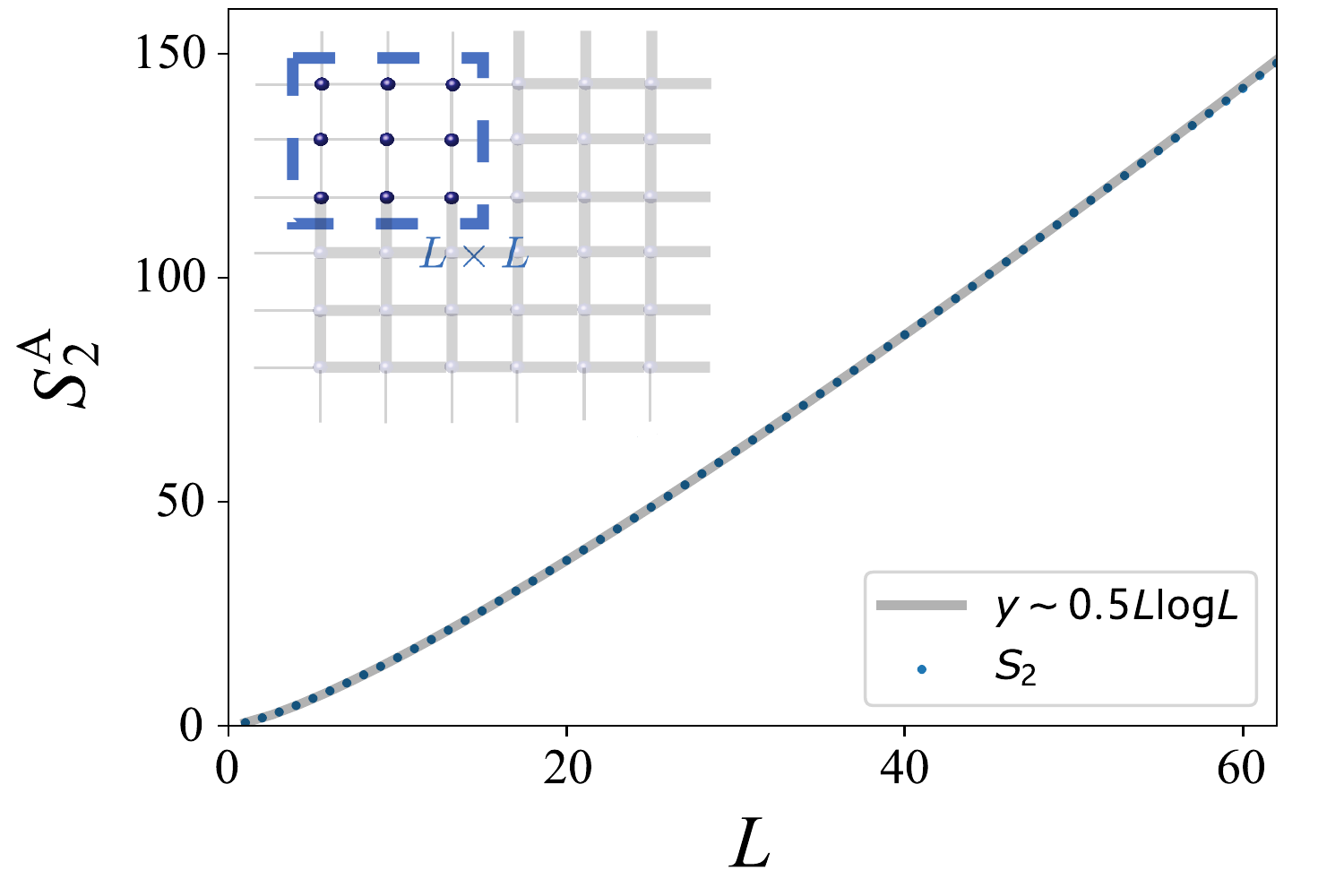}
	\caption{The EE of the free Fermi surface on square lattice Hubbard model. The $L\log L$ scaling behavior manifests with the coefficient computed from the Widom-Sobolev formula~\cite{gioev2006entanglement,
	leschkeScaling2014,sobolevSchatten2014,
	sobolevWiener2015,swingle2010entanglement,
	hellingSpecial2010,jiangFermion2022}. We choose system size $160\times 160$ to be closed to the thermodynamics limit, and $L$ is the length of considered square region. The grey line indicating Widom conjecture is guided by eyes. One expect $S_2 = 0.5 L \log L + O(L)$.}
	\label{fig:figS1}
\end{figure}  

\begin{figure}[htp!]
	\includegraphics[width=0.4\columnwidth]{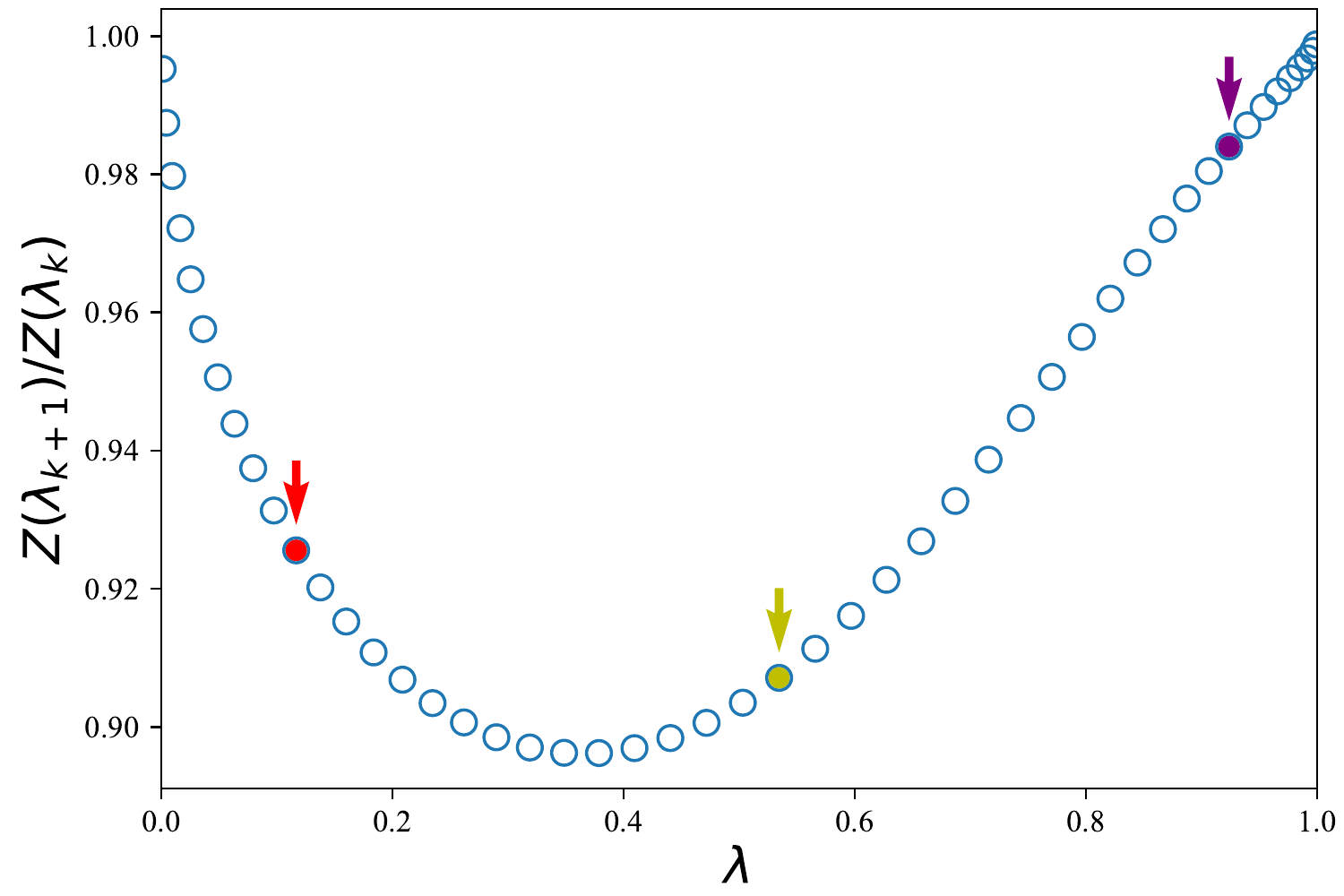}
	\includegraphics[width=0.4\columnwidth]{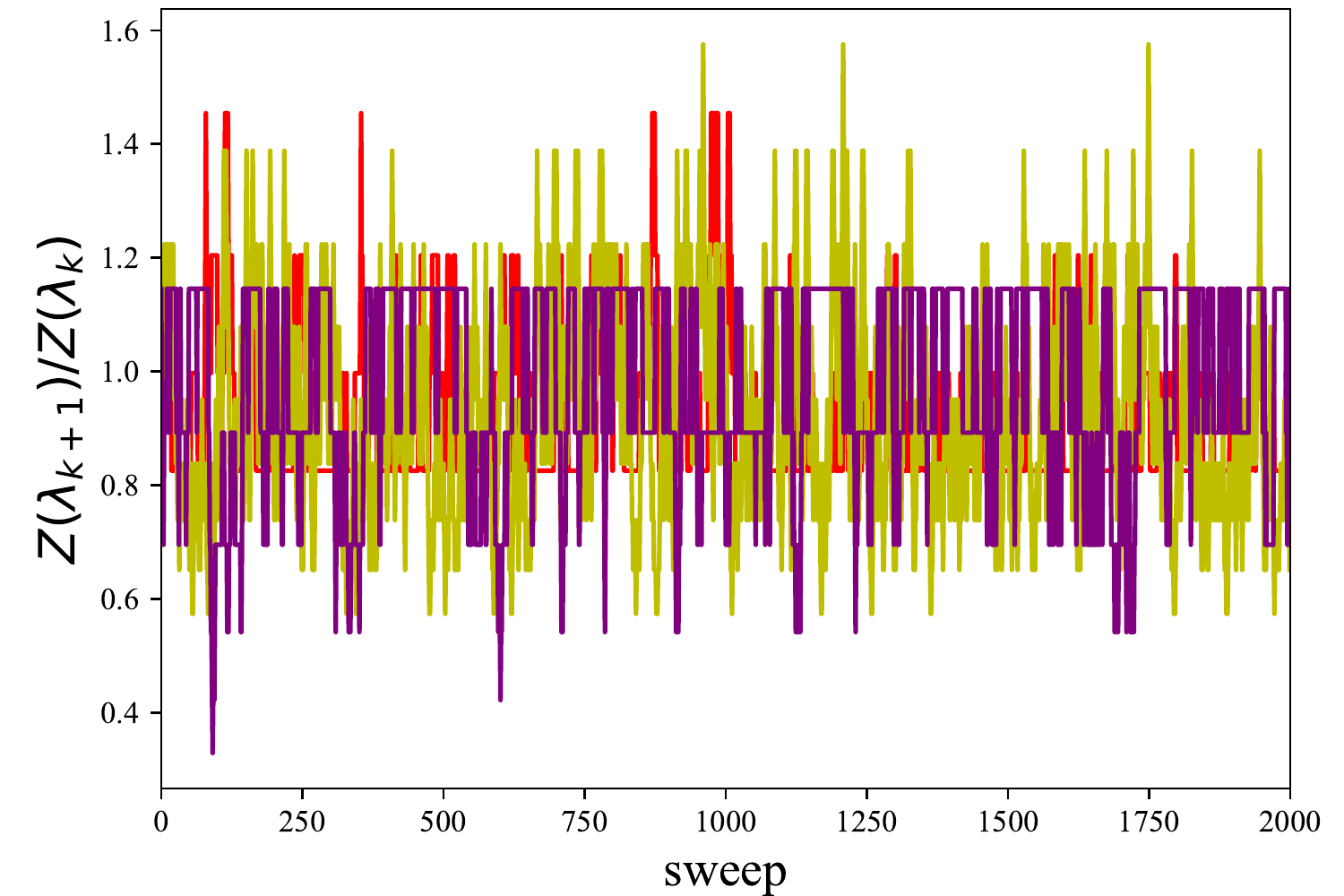}
	\caption{(a)$\frac{Z(\lambda_{k+1})}{Z(\lambda_k)}$ for $U=0, L=4$ by incremental method with $\lambda\in[0,1]$. As discussed in the main text, around Eqs.(4) and (5), each piece has the value of scale unity. Likewise in Fig. 3, the three arrows in panel(d) point out the position of three different $\lambda_k \sim 0.11,0.53,0.92 $ values where the time series are shown in panel(b). The observables are now normal distributed.}
	\label{fig:figS2}
\end{figure}  

We discuss the known results of EE in the free fermion limit $U=0$. We choose a square region with its side of length $L$, and the total system size $160\times 160$ to be closed to the thermodynamics limit. In such case, given the absence of the need for auxiliary field updates, we can directly compute the EE from Green's function matrices using Eq.(2) in the main text with $W^{s_1} = W^{s_2}=1$. The obtained results, as shown in Fig.~\ref{fig:figS1}, are consistent with analytical computation from the Widom-Sobolev formula~\cite{gioev2006entanglement,leschkeScaling2014,sobolevSchatten2014,sobolevWiener2015,swingle2010entanglement,hellingSpecial2010,jiangFermion2022}, where the coefficient of the $L\log L$ term precisely governs the data. According to the formula, one has  the following form of von Neumann entropy $S$,
\begin{equation}
	S^A(U=0) \sim  \frac{1}{12} \frac{L^{d-1}\log L}{(2\pi)^{d-1}} \int_{\partial \Omega} \int_{\partial\Gamma} |n_x \cdot n_p| \mathbf{d}S_x \mathbf{d}S_p
\label{eq:eqS4}
\end{equation}
where $\partial \Gamma$, $\partial \Omega$ are the boundaries of the Fermi sea and the region considered, $n_p$, $n_x$ are the unit normals to these boundaries. Note $\mathbf{d}S_x$ integrates on the box region with unit length, while $\mathbf{d}S_p$ on the momentum space. The integration can be regarded as the total length of projected Fermi surface on each side of the box. Since $U=0$, we have diamond Fermi surface, which contributes $4\pi$ for one side. Finally, we have $S^A \sim \frac{4\pi \times 4}{12} \frac{1}{2\pi} L \log L=\frac{2}{3} L \log L$. Note for free system, one has the relation between von Neumann entropy and the 2nd R\'enyi entropy $S^A_2 = \frac{3}{4} S^A$\cite{pasqualeExact2012}. Therefore, we expect $\frac{1}{2} L \log L$ leading term for the R\'enyi EE in Fig.~\ref{fig:figS1}. The total expression for this free limit is given in Eq.(3) where $a=\frac{1}{2}$ serves as the leading term coefficient determined by the region and Fermi surface.  As shown in the grey line in Fig.~\ref{fig:figS1}, the $\frac{1}{2} L \log L$ curve indeed goes through the data points.

In addition, as the Fig. 3 (d) in the main text, we show the ratio $\frac{Z(\lambda_{k+1})}{Z(\lambda_k)}$ with $\lambda_k\in[0,1]$ at the $U=0$ limit, which could be exactly computed for small system size, e.g. $L=4$. We divided $\lambda$ from 0 to 1 into 50 equal slices, and plot the ratio for adjacent two $\lambda$s in Fig.~\ref{fig:figS2} (a). As we expect, the ratio is closed to 1. To carefully study the distribution of the new observables for the incremental method, we plot the histogram at several $\lambda$ in Fig.~\ref{fig:figS2} (b). We find the distribution is almost a peak closed to O(1), thus avoid the sampling problem of the direct method. 

\section{Convergence of R\'enyi entanglement entropy $S_2^A$}

As shown in the manuscript, we convert the computation of $e^{-S_2^A}$ into a parallel execution of incremental ratios as
 \begin{equation}
 	e^{-S_2^A}=\frac{Z(1)}{Z(0)}:=\frac{Z(\lambda_1)}{Z(0)}\frac{Z(\lambda_2)}{Z(\lambda_1)}\cdots\frac{Z(\lambda_{k+1})}{Z(\lambda_k)} \cdots\frac{Z(1)}{Z(\lambda_{N_\lambda})},
 	\label{eq:eqs8}
 \end{equation}
  where $Z(\lambda)=\sum_{C \subseteq A} \lambda^{N_C}(1-\lambda)^{N_A-N_C} Z_C$~\cite{dEmidioEntanglement2020,dEmidioUniversal2022} with $\lambda \in[0,1]$, $N_C$ ($N_A$) is the number of site in region $C$ $(A)$ and $Z_C$ is partition function with entanglement region $C$. 
 
\begin{figure}[htp!]
	\includegraphics[width=0.5\columnwidth]{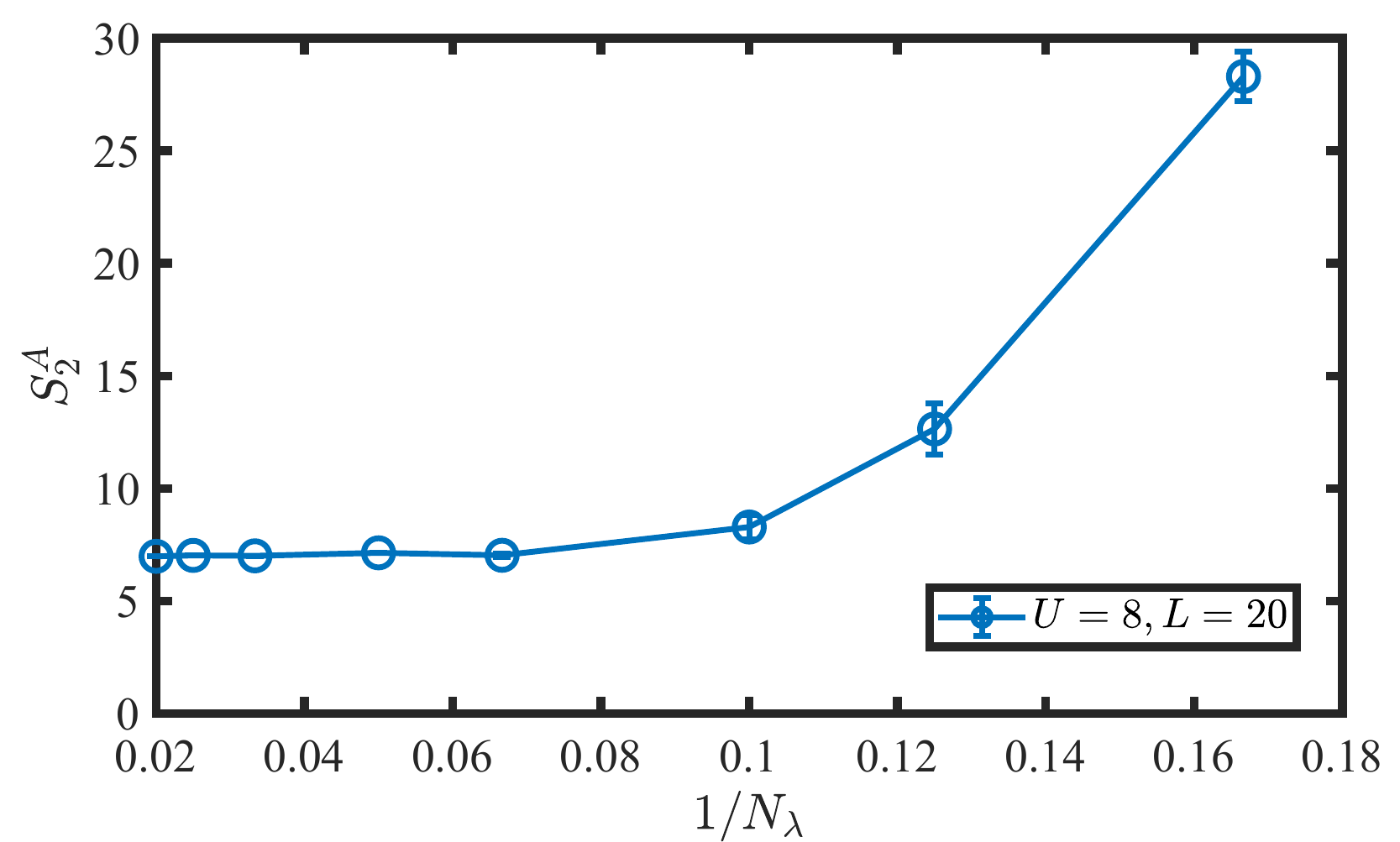}
	\caption{The convergence value of 2nd R\'enyi entanglement entropy $S_2^A$ respect to $N_\lambda=6,8,10,15,20,30,40,50$ for $U=8,L=20$.  }
	\label{fig:figS3}
\end{figure}

 All data in the main text is with the setting $N_\lambda=50$, and now we test convergence of $S_2^A$ with respect to $N_{\lambda}$, where $L=20,U=8$ and $\lambda_k=\left[\sin\frac{(0.002+N_{\lambda}(k-1))\pi}{2}\right]^2$. As shown in the Fig.~\ref{fig:figS3}, we select different $N_\lambda=6,8,10,15,20,30,40,50$ to calculate 2nd R\'enyi entanglement entropy $S_2^A$, and it can be seen that when $N_\lambda$ increases, the entanglement entropy value $S_2^A$ gradually converges and the errorbar becomes smaller and smaller.
 
\begin{figure}[htp!]
	\includegraphics[width=0.5\columnwidth]{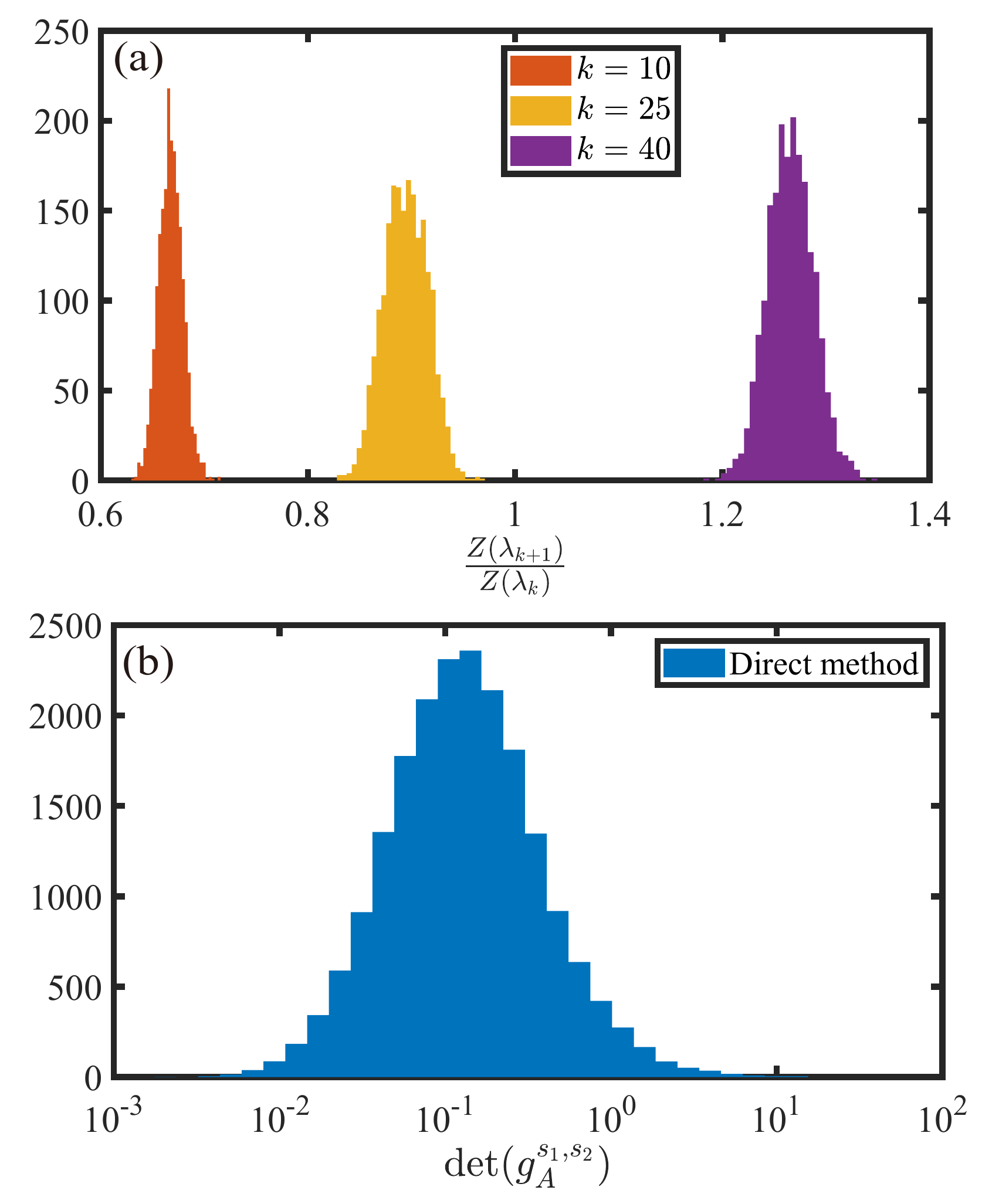}
	\caption{The histgram for incremental and direct methods with $U=8,L=4$. (a) The histgram of $\frac{Z(\lambda_{k+1})}{Z(\lambda_k)}$ for incremental method, narrow Gaussian distributions can be observed. (b) The histgram of  $\det(g_A^{s_1,s_2})$ for direct method, a Gaussian distributions can only be observed on the log scale, implying a large variance in the direct method.}
	\label{fig:figS4}
\end{figure}

In addition, in order to make connection with the Fig 3(b) and (c) in the main text, we also show the convergence of the Monte Carlo data itself here. As shown in the Fig.~\ref{fig:figS4}(a) , we draw the histogram of the data  $\frac{Z(\lambda_{k+1})}{Z(\lambda_k)}$, which correspond to Fig 3(c). And the histogram of the $\det(g_A^{s_1,s_2})$ correspond to Fig 3(b). Note that for the incremental method, the data approximates a good Gaussian distribution, while for the direct method, the data is a narrow Gaussian distribution on the log scale, which means that the variance of the original data is very large, which is one of the reasons why the direct method may not be easy to get accurate $S_2^A$. The parameters are $U=8,L=4$, which is alse same as that in Fig 3.

\end{document}